\documentclass[aps,prd,reprint,amsmath,amssymb,superscriptaddress,  nofootinbib,longbibliography,floatfix]{revtex4-2}

\usepackage[T1]{fontenc}
\usepackage[utf8]{inputenc}
\usepackage{amsmath,amssymb,mathtools}
\usepackage{bm}
\usepackage{microtype}
\usepackage[hidelinks]{hyperref}
\usepackage[normalem]{ulem}

\newcommand{\R}{\mathbb{R}}
\newcommand{\dd}{\mathrm{d}}
\newcommand{\ii}{\mathrm{i}}
\newcommand{\ee}{\mathrm{e}}
\newcommand{\1}{\mathbb{I}}
\newcommand{\kb}{{\bf k}_{\perp}}
\newcommand{\xb}{{\bf x}_{\perp}}
\newcommand{\xpb}{{\bf x}'_{\perp}}
\newcommand{\pM}{\partial M}

\hypersetup{
    colorlinks=true,
    citecolor=blue,
    linkcolor=blue,
    urlcolor=blue,
}

\newtheorem{proposition}{Proposition}

\begin{document}

\title{Robin–Hadamard states for unstable field in half-Minkowski spacetime}

\author{Bruno S. Felipe}
\email[]{felipe.bruno@ufabc.edu.br}
\affiliation{Centro de Matemática, Computação e Cognição, Universidade Federal do ABC (UFABC), 09210-170 Santo André, São Paulo, Brazil}

\author{Jo\~ao P. M. Pitelli}
\email[]{pitelli@unicamp.br}
\affiliation{Departamento de Matem\'atica Aplicada, Universidade Estadual de Campinas, 13083-859 Campinas, S\~ao Paulo, Brazil}

\author{Claudio Dappiaggi}
\email[]{claudio.dappiaggi@unipv.it}
\affiliation{Dipartimento di Fisica ``Alessandro Volta'', Universit\`a degli Studi di Pavia \& INFN, Sezione di Pavia \& INdAM, Sezione di Pavia, Via Bassi 6, I-27100 Pavia, Italia}


\begin{abstract}

We investigate the quantization of a real massless scalar field in four-dimensional half-Minkowski spacetime subject to an attractive Robin boundary condition. We show that the unstable modes, required for classical spectral completeness, are also essential to recover the canonical commutation relations (CCR) of the complete quantum field. By specifying the Cauchy covariances of the unstable sector, we construct a family of positive quasifree states compatible with the conservation of total energy. Within this family, positivity is incompatible with stationarity, so that the resulting states necessarily break the time-translation symmetry of the background. We also obtain a closed position-space representation of the two-point function for the chosen stable vacuum sector, including the contribution fixed by the CCR in the unstable sector. The resulting expression exhibits the direct and reflected singularities, including Robin-dependent logarithmic terms. Since the remaining unstable contribution is smooth, all states in the constructed family share the same singularity structure, consistent with the Robin--Hadamard framework.
\end{abstract}

\maketitle

\section{Introduction}
\label{sec:introduction}

In quantum field theory in curved spacetime, the presence of a timelike boundary fundamentally alters the quantization scheme. Since the field can interact causally with the boundary, the wave equation alone is insufficient to uniquely determine its dynamics without imposing suitable boundary conditions~\cite{GinouxMurro2022}. For static but non-globally hyperbolic backgrounds, such conditions can be formulated in terms of self-adjoint extensions of the spatial part of the wave operator, with each admissible extension determining the field evolution and its associated conserved energy~\cite{Wald1980,IshibashiWaldII}. In particular, when separation of variables reduces the boundary problem to a one-dimensional Sturm--Liouville operator with a single regular endpoint requiring boundary data, its self-adjoint extensions are characterized by a one-parameter family of Robin-type boundary conditions, encompassing Dirichlet and Neumann as particular cases~\cite{de2023hidden,DappiaggiFerreiraMarta2018,Morley2020Quantum,PitelliEtAl2025}.

Robin boundary conditions are not merely a technical choice---by determining the field dynamics; they also influence the spectral properties of the spatial operator and, consequently, the structure of the quantum theory. In particular, for an attractive Robin parameter, the spatial operator may develop a negative spectral sector associated with boundary-localized modes whose frequencies become purely imaginary~\cite{PhysRevD.89.025020,Asorey2015Edge,DappiaggiFerreira2016,FelipePitelli2024}. These modes exhibit exponential growth or decay in time, making the usual positive-frequency decomposition meaningless for the unstable sector and obstructing the standard construction of a stationary vacuum state. This raises the question of how such instabilities can be accommodated within a consistent quantum description. We therefore aim to investigate how to incorporate the unstable sector into the quantum theory and construct suitable quantum states on the full field algebra.

Following Ref.~\cite{KayWald1991}, we approach quantization by constructing {\it quasifree states}, characterized by positive two-point functions whose antisymmetric part is fixed by the canonical commutation relations (CCR). Within this framework, we show that attempting to circumvent the instability by discarding the unstable sector and retaining only the oscillatory contributions violates the completeness of the spectral decomposition. This omission has direct consequences for the field algebra, since the resulting antisymmetric two-point function fails to reproduce the equal-time normalization of the causal propagator and, consequently, the CCR. Moreover, this inconsistency cannot be resolved by modifying the symmetric part of the two-point function, since such modifications affect the choice of state without restoring the missing contribution to the causal propagator, which is fixed by the boundary value problem. A consistent quantization must therefore account for the complete spectral decomposition, including the unstable sector, and establish the conditions under which its dynamics admits positive quasifree states. We further show that, within the family of states constructed here, positivity is incompatible with stationarity.

In addition to positivity and compatibility with the CCR, a physically admissible state is expected to satisfy the Hadamard condition, which constrains the singularity structure of the two-point function and allows the definition of local observables, such as the field fluctuations and the renormalized stress-energy tensor, through the point-splitting technic~\cite{Radzikowski1996}. In the presence of a timelike boundary, however, the propagation of singularities must also account for null rays reflected at the boundary, as illustrated by the Robin wave propagators studied in Ref.~\cite{Bondurant_2005}. In this direction, Costeri et al.~\cite{CosteriEtAl2026} constructed a Hadamard parametrix compatible with Robin boundary conditions in half-Minkowski spacetime, subsequently extending this construction to a class of globally hyperbolic spacetimes with timelike boundaries~\cite{CosteriDappiaggiJuarezAubry2026}. Their analysis shows that the boundary-adapted singular structure contains both the standard direct Hadamard contribution and additional terms associated with reflected null rays, leading to the notion of {\it Robin--Hadamard} states.

The existing constructions, however, do not directly address the presence of an unstable spectral sector. In this way, our purpose is not to construct a new Hadamard parametrix, but rather to investigate whether the two-point function of a positive quasi-free state incorporating unstable modes can exhibit the Robin--Hadamard singularity structure. To this end, we construct the complete two-point function, including the unstable contribution, and examine its singularities in light of the results established by Costeri et al.~\cite{CosteriEtAl2026}. In particular, we identify the contributions associated with direct and reflected null propagation and investigate whether the unstable sector introduces any additional singularities. This allows us to examine the compatibility between boundary-induced instabilities, positivity, the CCR, and the Robin--Hadamard condition in a regime where the spatial operator is no longer non-negative.

To address these questions, we consider a real massless scalar field in a four-dimensional half-Minkowski spacetime, subject to an attractive Robin boundary condition. This background is described by the manifold with boundary $M=\{x=(t,\xb,z)\in\R^4:z\geq0\}$, equipped with the line element
\begin{equation}
 \dd s^2=-\dd t^2+\dd\xb^2+\dd z^2,
 \qquad z\geq0.
 \label{eq:metric}
\end{equation}
Here, $\xb=(x,y)\in\R^2$ denotes the spatial coordinates parallel to the boundary. The timelike boundary is located at $z=0$ and is given by $\pM={(t,\xb,0):(t,\xb)\in\R^3}$, while $M^\circ:=M\setminus\pM$ denotes the spacetime interior. This geometry allows us to investigate the spectral properties induced by Robin boundary conditions and their consequences for quantization without additional complications arising from spacetime curvature. Moreover, since $M^\circ$ is conformally related to the Poincaré patch of anti-de Sitter spacetime~\cite{BayonaBraga2007}, the present analysis also provides a useful starting point for investigating analogous boundary-induced instabilities in curved backgrounds. We leave such extensions for future work.

The remainder of this paper is organized as follows. In Sec.~\ref{sec:action}, we derive the field dynamics and the associated conserved energy, emphasizing the role of the attractive Robin parameter and its distinction from the regime considered in Ref.~\cite{CosteriEtAl2026}. We also establish the spectral decomposition of the spatial operator, identifying the stable and unstable sectors. In Sec.~\ref{sec:two-point-ccr}, we consider the integral representation of the complete two-point function, including the unstable contribution, and establish the regime in which the resulting function can be positive, compatible with the CCR, and stationary. In Sec.~\ref{sec:hadamard}, we obtain a closed position-space representation of the singular contributions and identify the direct and boundary-reflected singularities, including the Robin-dependent terms. We then examine these singularities in light of the Robin--Hadamard framework, identifying a state-independent subtraction term through the spectral representation. Finally, Sec.~\ref{sec:conclusion} summarizes our results and discusses their implications for the quantization of fields in the presence of instabilities.

\section{Robin dynamics from an action}
\label{sec:action}

Let $(M,\eta)$ be half-Minkowski spacetime with the line element~\eqref{eq:metric}. At the boundary $\pM$, we define the outward-pointing unit normal vector $n^{\mu}=-\delta_{z}^{\mu}$ so that $n^{\mu}\partial_{\mu}=-\partial_z$. Since the spacetime has a boundary, we follow Saharian's prescription~\cite{Saharian2004}, according to which the dynamics of a real scalar field is determined by the action
\begin{equation}
 S[\phi]
 =-\frac12\int_{M}\dd^{4}x\,
   \partial_{\mu}\phi\,\partial^{\mu}\phi
 +\frac{\beta}{2}\int_{\pM}\dd^{3}x\,\phi^{2}.
 \label{eq:action}
\end{equation}
Here, $\phi:M\to\R$ is a massless field coupled to the wall through the finite Robin boundary parameter $\beta\in\R$, which has dimensions of inverse length. Unless we are explicitly comparing different boundary conditions, we take $\beta>0$ throughout this work.

Extremizing Eq.~\eqref{eq:action} with respect to $\phi$ and integrating by parts, we obtain
\begin{equation}
 \delta S={}\int_{M}\dd^{4}x\,
 (\partial^{\mu}\partial_{\mu}\phi)\,\delta\phi
 +\int_{\pM}\dd^{3}x\,
 \left(\partial_z\phi+\beta\phi\right)\delta\phi .
 \label{eq:variation}
\end{equation}
Stationarity under arbitrary variations therefore yields the wave equation
\begin{equation}
  \square_x\phi=\partial^\mu\partial_\mu \phi=0,
  \label{eq:wave}
\end{equation}
together with the regular Robin boundary condition
\begin{equation}
  \left.(\partial_z\phi+\beta\phi)\right|_{z=0}=0.
  \label{eq:robin}
\end{equation}
In this notation, $\beta=0$ gives the Neumann boundary condition, while the Dirichlet boundary condition is obtained as $|\beta|\to\infty$.

Varying Eq.~\eqref{eq:action} with respect to the metric, gives a classical stress-energy tensor with a distributional surface contribution~\cite{FelipePitelli2024}
\begin{equation}
 T_{\mu\nu}
 =\partial_{\mu}\phi\,\partial_{\nu}\phi
 -\frac12\eta_{\mu\nu}
   \partial_{\rho}\phi\,\partial^{\rho}\phi
 +\frac{\beta}{2}\,h_{\mu\nu}\phi^{2}\delta_{\pM},
 \label{eq:stress-total}
\end{equation}
where $h_{\mu\nu}=\eta_{\mu\nu}-n_{\mu}n_{\nu}$ is the metric induced on $\pM$. The boundary delta distribution is normalized by
\begin{equation}
 \int_M\dd^4 x\,\delta_{\pM}f
 =\int_{\pM}\dd^{3}x\,f,
\end{equation}
for every test function $f$. Thus, the Robin boundary condition contributes explicitly to the stress-energy tensor through a term localized at $\pM$. We stress that, in this work, we do not investigate the quantum counterpart of $T_{\mu\nu}$ which would require an investigation of the conformal anomaly in the presence of Robin boundary conditions.

Integrating the $tt$-component of the stress tensor \eqref{eq:stress-total} over a constant-time hypersurface yields the Hamiltonian
\begin{equation}
\begin{aligned}
 H={}&\frac{1}{2}\int\dd^{2}{\xb}
 \int_{0}^{\infty}\dd z\,
 \left[
   \dot{\phi}^{2}+(\partial_x\phi)^2+(\partial_y\phi)^2
   +(\partial_z\phi)^{2}
 \right]\\
 &-\frac{\beta}{2}\int\dd^{2}{\xb}\,\phi(t,{\xb},0)^{2}
 \label{eq:energy} 
\end{aligned}
\end{equation}
which generates the field dynamics subject to the Robin boundary condition. The term proportional to $\beta$ represents a quadratic potential localized at the boundary, whose sign determines whether the boundary interaction is attractive or repulsive. With the convention adopted in Eq.~\eqref{eq:robin}, we refer to $\beta>0$ as the attractive Robin sign, since it supports a bound mode localized in the direction normal to the boundary. For $\beta<0$, as considered in Ref.~\cite{CosteriEtAl2026}, the boundary contribution to the Hamiltonian is positive. On the other hand, for $\beta>0$, this contribution is negative, allowing the Hamiltonian to become indefinite.

Despite the absence of a positive-definite energy, the dynamics remain conservative with respect to energy. Indeed, taking the time derivative of Eq.~\eqref{eq:energy}, using the wave equation~\eqref{eq:wave}, and integrating by parts, we obtain
\begin{equation}
 \frac{\dd H}{\dd t}
 =-\int\dd^{2}\xb\,
 \dot\phi(\partial_z\phi+\beta\phi)|_{z=0}=0,
 \label{eq:conservation}
\end{equation}
where we have assumed sufficient decay at spatial infinity. The final equality follows directly from the Robin boundary condition~\eqref{eq:robin}. Thus, the bulk and boundary contributions together define a conserved total energy, even though it is not positive definite for the attractive Robin sign. As we shall see, restricting the field to the nonnegative spectral subspace of the spatial operator would not resolve this issue without altering the theory, since it would remove modes required for the completeness of the spectral decomposition.

\subsection{Spectrum and mode solutions}
\label{subsec:spectrum}

The wave equation~\eqref{eq:wave}, together with the Robin boundary condition, can be expressed as
\begin{equation} \label{eq:spatial-operator}
\ddot\phi+\Delta_\beta\phi=0,
\end{equation}
where $\Delta_\beta=-\partial_x^2$ $-\partial_y^2-\partial^2_z$ is the spatial operator whose domain is 
\begin{equation}
\begin{aligned}
D(\Delta_\beta):=&\{u\equiv u(x,y,z)=\sum\limits_{i=1}^Nf_i(x,y)g_i(z)\;|\; N\in\mathbb{N} \\
& \mathrm{and}\;f_i\in H^2(\mathbb{R}^2)\;\textrm{while}\;g_i\in D_\beta(\partial^2_z)\},
\end{aligned}
\end{equation}
where $H^2(\mathbb{R}^2)$ is the second Sobolev space on $\mathbb{R}^2$, while, denoting by $\mathbb{R}_+:=[0,\infty)$,
$$D_\beta(\partial^2_z):=\{g\in H^{2}(\R_{+})\; |\; g'(0)+\beta g(0)=0\}.$$
Observe that, by Sobolev lemma, every function lying in $H^2(\R_+)$ is also differentiable and, hence evaluating it together with its derivative at $z=0$ is meaningful. Notice that $D(\Delta_{\beta})$ is a core for the operator $\Delta_\beta$ on $L^2(\R^2\times \R_+)$ and, hence, $\Delta_\beta$ is thereon essentially self-adjoint. Its unique self-adjoint extension is referred to as the Laplacian encoding Robin boundary conditions.

In order to construct a solution of Equation \eqref{eq:wave} with Robin boundary conditions as per Equation \eqref{eq:robin}, by spectral calculus we can look for $f_\lambda\in D_\beta(\partial^2_z)$ such that
\begin{equation}
 \partial^2_z f_\lambda=\lambda f_\lambda,
 \label{eq:normal-eigenvalue}
\end{equation}
where the eigenvalue $\lambda$ can be positive or negative, since the Robin extension need not be non-negative. To simplify the notation, henceforth, we denote by $A_\beta$ the self-adjoint extension of $\partial^2_z$ on $D_\beta(\partial^2_z)$, which is nothing but its graph closure. Under these premises, we have the quadratic form
\begin{equation}
 \langle f,A_\beta f\rangle
 =\int_0^\infty\dd z\,|f'(z)|^2
 -\beta|f(0)|^2,
 \label{eq:normal-quadratic-form}
\end{equation}
which can be negative for $\beta>0$. Notice that the negative boundary term in Eq.~\eqref{eq:normal-quadratic-form} corresponds precisely to the boundary potential appearing in the Hamiltonian~\eqref{eq:energy}. Thus, the attractive Robin condition allows the normal operator to develop a negative spectral sector, which, as a consequence, renders the total field energy indefinite. Nevertheless, the existence of negative spectral values does not imply that the spatial operator is unbounded below. Completing the square in Eq.~\eqref{eq:normal-quadratic-form}, we obtain
\begin{equation}
 \langle f,A_\beta f\rangle
 =\|f'+\beta f\|^2-\beta^2\|f\|^2
 \geq-\beta^2\|f\|^2.
 \label{eq:semibounded-normal-operator}
\end{equation}
showing that $A_\beta$ is bounded below by $-\beta^2$. Since the tangential Laplacian $-\partial_x^2-\partial_y^2$ is non-negative, the same lower bound holds for the full spatial operator $\Delta_\beta$.

To solve Eq.~\eqref{eq:normal-eigenvalue}, we first consider the positive spectral values $\lambda=q^2>0$, which belong to the continuous spectrum of $A_\beta$. The corresponding generalized eigenfunctions can be chosen as
\begin{equation}
 \rho_q(z)
 =\sqrt{\frac{2}{\pi}}\,
 \frac{q\cos(qz)-\beta\sin(qz)}
 {\sqrt{q^{2}+\beta^{2}}},
 \label{eq:rhoq}
\end{equation}
which satisfy $A_\beta\rho_q=q^2\rho_q$. In addition, $A_{\beta}$ also admits a discrete negative eigenvalue, $\lambda=-\beta^2$, corresponding to the normalized eigenfunction
\begin{equation}
 \psi_b(z)=\sqrt{2\beta}\,\ee^{-\beta z},
 \label{eq:psib}
\end{equation}
with $A_\beta\psi_b=-\beta^2\psi_b$ and $|\psi_b|=1$. Since $\psi_b$ decays exponentially away from the boundary, it represents a bound state localized in the normal direction by the attractive Robin condition. Notice that this eigenvalue also saturates the lower bound established in Eq.~\eqref{eq:semibounded-normal-operator}.

The self-adjointness of $A_\beta$ ensures that its distinct spectral subspaces are mutually orthogonal. With the normalization adopted above, the generalized eigenfunctions satisfy
\begin{align}
 \int_{0}^{\infty}\dd z\,\rho_q(z)\rho_{q'}(z)
 &=\delta(q-q'),
 \label{eq:rho-orthogonality}\\
 \int_{0}^{\infty}\dd z\,\rho_q(z)\psi_b(z)
 &=0.
 \label{eq:rho-bound-orthogonality}
\end{align}
Together, the generalized continuum eigenfunctions and the normal bound state provide the resolution of the identity
\begin{equation}
 \int_{0}^{\infty}\dd q\,\rho_q(z)\rho_q(z')+\psi_b(z)\psi_b(z')=\delta(z-z').
 \label{eq:normal-completeness}
\end{equation}
Here, the integral over $q$ represents the kernel of the projector onto the continuous spectral subspace, while the second term corresponds to the rank-one projector onto the bound state. Their sum reconstructs the identity operator, showing that both contributions are necessary for the completeness of the spectral decomposition.

The spectrum of the normal operator is therefore
\begin{equation}
\sigma(A_\beta)={-\beta^2}\cup[0,\infty),\qquad \beta>0.
\label{eq:normal-spectrum}
\end{equation}
For $\beta<0$, as considered in Ref.~\cite{CosteriEtAl2026}, the formal bound-state solution $\ee^{-\beta z}$ grows exponentially as $z\to\infty$ and is not square-integrable. Consequently, the discrete negative eigenvalue is absent, and the spectrum reduces to $[0,\infty)$. The same spectrum is obtained for Dirichlet and Neumann boundary conditions, neither of which supports a normalizable bound state.

To construct the field solutions from the spectral decomposition obtained above, we first introduce the Klein--Gordon product on a constant-time hypersurface $\Sigma_0$, defined by
\begin{equation}
 (f,g)
 =\left.\ii\int\dd^{2}\xb\int_0^\infty\dd z\,
 \left(f^{\ast}\partial_tg-(\partial_t f^\ast)g\right)\right|_{t=0}.
 \label{eq:KG-product}
\end{equation}
For solutions satisfying the Robin condition, the Klein--Gordon current has no flux through the timelike boundary, and hence, Eq.~\eqref{eq:KG-product} is independent of the constant-time hypersurface on which it is evaluated~\cite{Morley2023Renormalized}.

Now, let $\kb=(k_x,k_y)$ denote the momentum parallel to the boundary, and write $k=|\kb|=\sqrt{k_x^2+k_y^2}$.  For the continuous spectrum of $A_\beta$, characterized by the normal wave number $q>0$, the full spatial operator $\Delta_\beta$ has spectral value $k^2+q^2$. Consequently, the corresponding solutions oscillate with frequency $\omega_{kq}=\sqrt{k^2+q^2}$, and the positive-frequency modes normalized with respect to Eq.~\eqref{eq:KG-product} can be written as
\begin{equation}
 u_{\kb q}(x)
 =\frac{\rho_q(z)}{2\pi\sqrt{2\omega_{k q}}}
 \ee^{-\ii\omega_{k q}t+\ii\kb\cdot\xb}.
 \label{eq:continuum-modes}
\end{equation}

The situation is different for the normal bound state $\psi_b$, whose contribution to the full spatial spectrum is given by $k^2-\beta^2$. Although the normal eigenvalue is negative, the resulting field solutions are not necessarily unstable. Indeed, for $k>\beta$, this spectral value remains positive, and the corresponding frequency $\Omega_k=\sqrt{k^2-\beta^2}$ is real. In this regime, the normalized positive-frequency modes are
\begin{equation}
 v_{\kb}(x)
 =\frac{\psi_b(z)}
 {2\pi\sqrt{2\Omega_{k}}}
 \ee^{-\ii\Omega_{k}t+\ii\kb\cdot\xb}.
 \label{eq:stable-bound-modes}
\end{equation}
For $k<\beta$, however, the spatial spectral value becomes negative, giving purely imaginary frequencies $\omega_k=\pm\ii\gamma_k$, where $\gamma_k=\sqrt{\beta^2-k^2}$. In this case, the temporal evolution is exponential rather than oscillatory, and a convenient pair of solutions is
\begin{equation}
 v_{\kb}^{\pm}(x)
 =\frac{\psi_b(z)}{2\pi\sqrt{2\gamma_{k}}}
 \ee^{\mp\gamma_{k}t+\ii\kb\cdot\xb}.
 \label{eq:unstable-bound-modes}
\end{equation}
Here, both solutions carry the same tangential Fourier phase, while their time dependence distinguishes the two directions of evolution: $v_{\kb}^{+}$ decays toward the future and grows toward the past, whereas $v_{\kb}^{-}$ exhibits the opposite behavior. Together, they span the unstable sector.

Unlike the oscillatory modes, the solutions~\eqref{eq:unstable-bound-modes} cannot be classified through the usual positive- and negative-frequency decomposition. Their Klein--Gordon products instead satisfy
\begin{equation}
\begin{aligned}
\left(v_{\kb}^{+},v_{\kb'}^{+}\right)
 &={}\left(v_{\kb}^{-},v_{\kb'}^{-}\right)=0,\\
 \left(v_{\kb}^{\pm},v_{\kb'}^{\mp}\right)
 &={}\pm\ii\delta^{2}(\kb-\kb').
\end{aligned}
\label{eq:unstable-KG-products}
\end{equation}
Thus, each member of the unstable pair has vanishing Klein--Gordon norm, while their mixed products are nonzero and purely imaginary. This structure will play an important role in the construction of the two-point function, since the unstable sector cannot be treated through the standard positive-frequency prescription.

At the threshold $k=\beta$, the temporal equation admits solutions with affine time dependence. Since the threshold circle has measure zero in the two-dimensional tangential momentum space, we do not include additional distributions supported on it.

Having identified the continuous, stable bound, and unstable bound contributions, we can now express a real classical solution with smooth, compatible Cauchy data and sufficient spatial decay as
\begin{equation}
    \begin{aligned}
        \phi(x)={}&\int_{\mathbb{R}^{2}}\dd^{2}\kb\int_0^\infty\dd q\,\left[a_{\kb q}u_{\kb q}(x)+\mathrm{c.c.}\right]\\ &+\int_{k>\beta}\dd^{2}\kb\,\left[b_{\kb}v_{\kb}(x)+\mathrm{c.c.}\right]\\ &+\int_{k<\beta}\dd^{2}\kb\,\left[c_{\kb}^{+}v_{\kb}^{+}(x)+c_{\kb}^{-}v_{\kb}^{-}(x)\right].
    \label{eq:total-field}
    \end{aligned}
\end{equation}
Here, $a_{\kb q}$ and $b_{\kb}$ are complex amplitudes associated with the oscillatory sectors, while $c_{\kb}^{\pm}$ determine the exponentially evolving contributions. All coefficients are fixed by the initial Cauchy data. Since the field is real, the unstable-sector coefficients must satisfy $(c_{\kb}^{\pm})^{*}=c_{-\kb}^{\pm}$.

Substituting Eq.~\eqref{eq:total-field} into Eq.~\eqref{eq:energy} and using the orthogonality of the spatial modes, we get
\begin{equation}
 \begin{aligned}
     H={}&\int_{\mathbb{R}^{2}} \dd^{2}{\kb}\int_0^\infty \dd q\,\omega_{kq}|a_{\kb q}|^2\\
     &+\int_{k>\beta}\dd^{2}{\kb}\,\Omega_k|b_{\kb}|^2-\int_{k<\beta}\dd^{2}{\kb}\,\gamma_k c_{\kb}^{+}c_{-\kb}^{-}.
 \end{aligned}
 \label{eq:classical-mode-hamiltonian}
\end{equation}
The first two terms correspond to the continuous and stable bound sectors, respectively, and have the usual positive harmonic-oscillator form. The last term, however, arises from the unstable sector and is not positive definite. By the reality conditions on $c_{\kb}^{\pm}$ and the symmetry under $\kb\mapsto-\kb$, its integral is real. Moreover, its bilinear structure is characteristic of the Hamiltonian of an inverted harmonic oscillator, whose energy is unbounded below~\cite{FelipePitelli2024,Lima2013}. Thus, the spectral decomposition makes explicit how the attractive Robin boundary condition gives rise to exponentially evolving modes while preserving the conservation of the total Hamiltonian.

\section{Two-point function and CCR}
\label{sec:two-point-ccr}

Although the background $(M,\eta)$ is static, the presence of an unstable sector prevents the usual construction of a stationary vacuum through a positive-frequency decomposition of the complete field~[see Eq.~\eqref{eq:total-field}]. We therefore adopt the algebraic formulation of quantum field theory~\cite{KayWald1991} (see also Refs.~\cite{Khavkine2015,Kay2006Quantum,Hack2016AlgebraicQF} for broader discussions), in which the field algebra is determined by the dynamics and the CCR, while centered positive quasifree states are characterized by the total two-point distributions without assigning a particular particle interpretation. 

Let $\widehat{\phi}$ denote the real quantum field operator. A centered quasifree state $\omega$ is completely characterized by its two-point distribution, defined for complex test functions $f,g\in C_0^\infty(M^\circ;\mathbb C)$ as
\begin{equation}
 W(f,g):=\omega\!\left(\widehat\phi(f)\widehat\phi(g)\right).
 \label{eq:two-point-definition}
\end{equation}
The corresponding distributional kernel $W\in\mathcal D'(M^\circ\times M^\circ)$ must be a bisolution of the field equation, $\Box_x W=\Box_{x'}W=0$, and satisfy the Robin boundary condition~\eqref{eq:robin} in each argument. Moreover, the Hermiticity of the field and positivity of the state require
\begin{equation}
 W(x,x')^*=W(x',x), \qquad W(f^*,f)\geq0,
\end{equation}
where the first relation is understood in the distributional sense.

These conditions must also be compatible with the CCR, which fix the antisymmetric part of the two-point function according to
\begin{equation}
 W(x,x')-W(x',x)=\ii E(x,x'),
 \label{eq:antisymmetric-E}
\end{equation}
where $E$ denotes the retarded-minus-advanced causal propagator associated. With our sign convention, its equal-time Cauchy data satisfy
\begin{equation}
\begin{aligned}
 \left.E(x,x')\right|_{t'=t}&=0,\\
 \left.\partial_{t'}E(x,x')\right|_{t'=t}
 &=\delta^{3}({\bf x}-{\bf x}').
 \label{eq:E-initial-data}
\end{aligned}
\end{equation}
Therefore, Eqs.~\eqref{eq:antisymmetric-E} and \eqref{eq:E-initial-data} are the two-point function formulation of the CCR.

To distinguish the part determined by the field dynamics from the freedom associated with the choice of state, we introduce the symmetric part of the two-point function
\begin{equation}
 S(x,x'):=\frac12\left[W(x,x')+W(x',x)\right],
 \label{eq:W-symmetric-decomposition}
\end{equation}
so that
\begin{equation}
  W(x,x')=S(x,x')+\frac{\ii}{2}E(x,x').
\end{equation}
This construction makes the division of roles explicit: $E$ is fixed by the dynamics and the CCR, whereas the real symmetric bisolution $S$ contains the state dependence and is constrained by positivity. We now turn to the construction of these contributions for the Robin problem in the presence of unstable modes.

\subsection{Spectral completion of the CCR}
\label{subsec:continuum}

On the stable spectral subspaces, positive-frequency modes give the continuous and stable bound contributions to the two-point function:
\begin{align}
 W_c(x,x')&=\int_{\R^{2}}\dd^{2}\kb\int_0^\infty\dd q\,
 u_{\kb q}(x)u_{\kb q}^*(x'),
 \label{eq:continuous-wightman}\\
 W_b(x,x')&=\int_{k>\beta}\dd^{2}\kb\,
 v_{\kb}(x)v_{\kb}^*(x').
 \label{eq:stable-bound-wightman}
\end{align}
Accordingly, we write the total stable contribution as
\begin{equation}
 W_{\mathrm{st}}=W_c+W_b.
 \label{eq:stable-wightman}
\end{equation}
These mode integrals are understood as distributional Wightman boundary values, obtained by introducing convergence factors $\ee^{-\epsilon\omega_{kq}}$ and $\ee^{-\epsilon\Omega_k}$, respectively, and then taking $\epsilon\downarrow 0$. They define stationary positive-type kernels on their respective spectral subspaces. Moreover, the threshold behavior of $W_b$ does not introduce a divergence, since $\Omega_k^{-1}=\mathcal{O}((k-\beta)^{-1/2})$ as $k\downarrow\beta$, which is locally integrable. Similarly, for fixed $\beta>0$, the continuum eigenfunctions satisfy $\rho_q(z)=\mathcal{O}(q)$ as $q\to0$ on compact normal intervals, ensuring that $W_c$ has no massless infrared divergence in four dimensions.

For the unstable sector, no analogous positive-frequency construction is available. Its contribution to the antisymmetric part can nevertheless be obtained by antisymmetrizing the products $v_{\kb}^{+}(x)(v_{\kb}^{-}(x'))^*$ and $v_{\kb}^{-}(x)(v_{\kb}^{+}(x'))^*$. Together with the stable-sector contribution, this yields the complete causal propagator:
\begin{widetext}
\begin{equation}
\begin{aligned}
 E(x,x')={}&-\int_{\mathbb{R}^{2}}\frac{\dd^{2} \kb}{(2\pi)^{2}}\ee^{\ii\kb\cdot(\xb-\xpb)}
 \int_0^\infty \dd q\,\rho_q(z)\rho_q(z')
 \frac{\sin[\omega_{k q}(t-t')]}{\omega_{k q}}
 \\
&-\psi_b(z)\psi_b(z')\left[\int_{k>\beta}
 \frac{\dd^{2}{\kb}}{(2\pi)^{2}}\ee^{\ii\kb\cdot(\xb-\xpb)}\frac{\sin[\Omega_{k}(t-t')]}{\Omega_k}
 +\int_{k<\beta}
 \frac{\dd^{2}{\kb}}{(2\pi)^{2}}\ee^{\ii\kb\cdot(\xb-\xpb)}\frac{\sinh[\gamma_{k}(t-t')]}{\gamma_k}
 \right].
\end{aligned}
\label{eq:full-E}
\end{equation}
\end{widetext}
Each contribution in Eq.~\eqref{eq:full-E} is a bisolution of the wave equation satisfying the Robin boundary condition. Moreover, all terms vanish at equal times, so the first Cauchy datum in Eq.~\eqref{eq:E-initial-data} is automatically satisfied.

To verify the second datum, we first identify how the stable and unstable momentum ranges contribute to the bound-state projector. For this purpose, let us define the tangential kernels
\begin{equation}
 \Pi_{\gtrless}(\xb-\xpb)
 :=\int_{k\gtrless\beta}
 \frac{\dd^{2}\kb}{(2\pi)^{2}}
 \ee^{\ii\kb\cdot(\xb-\xpb)}.
 \label{eq:tangential-projectors}
\end{equation}
The threshold $k=\beta$ has measure zero in the two-dimensional tangential momentum space, so
\begin{equation}
 \Pi_>+\Pi_<
 =\delta^{2}(\xb-\xpb).
 \label{eq:tangential-completeness}
\end{equation}
Thus, although the bound modes have different temporal behaviors in the two momentum regions, both contributions are necessary to reconstruct the complete bound-state projector.

 Differentiating $E$ with respect to $t'$, and evaluating at $t'=t$, we obtain
\begin{equation}
  \begin{aligned}
    &\left.\partial_{t'}E(x,x')\right|_{t'=t}\\
    &=\delta^{2}({\xb}-{\xpb})\left[\int_0^\infty\dd q\, \rho_q(z)\rho_q(z')+\psi_b(z)\psi_b(z')\right],
  \end{aligned}
\end{equation}
where we have used Eq.~\eqref{eq:tangential-completeness}. The expression in square brackets is precisely the resolution of the identity for the normal operator, Eq.~\eqref{eq:normal-completeness}. Consequently,
\begin{equation}
  \left.\partial_{t'}E(x,x')\right|_{t'=t}=\delta^{2}({\xb}-{\xpb})\delta (z-z'),
  \label{eq:completeness-ccr}
\end{equation}
which reproduces the second Cauchy datum in Eq.~\eqref{eq:E-initial-data}. Therefore, the complete spectral kernel satisfies both the field equation and the equal-time normalization required by the CCR.

As discussed in Sec.~\ref{subsec:spectrum}, the spatial operator $\Delta_\beta$ is self-adjoint and bounded below by $-\beta^2$, so the spectral theorem guarantees a unique, well-posed Cauchy evolution, even in the presence of negative spectral values. Moreover, the locality of the Robin boundary condition ensures finite propagation speed, so the complete spectral kernel, which satisfies the wave equation, the Robin boundary condition, and the prescribed Cauchy data, coincides with the retarded-minus-advanced causal propagator for the attractive Robin problem (see Theorem~30 of Ref.~\cite{Dappiaggi2019} for the existence and uniqueness of advanced and retarded fundamental solutions with causal support).

This result makes explicit why the unstable sector cannot be discarded from the quantum theory. The equal-time normalization of the causal propagator relies on the completeness of both the normal and tangential spectral decompositions. If the unstable contribution were omitted, the second Cauchy datum would lack the term
\begin{equation}
    \begin{aligned}
     &\psi_b(z)\psi_b(z')\Pi_<(\xb-\xpb)
     \\
     &\quad=\psi_b(z)\psi_b(z')
     \int_{k<\beta}\frac{\dd^{2}\kb}{(2\pi)^{2}}
     \ee^{\ii\kb\cdot(\xb-\xpb)}.
    \end{aligned}
 \label{eq:missing-unstable-projector}
\end{equation}
This missing projector is nonzero, and its absence would prevent the antisymmetric part of the two-point function from reproducing the CCR. Moreover, the inconsistency cannot be remedied by modifying the symmetric part of $W$, since such modifications affect the choice of quantum state without restoring the missing contribution to the causal propagator. We therefore conclude that the unstable spectral sector is indispensable for the canonical completeness of the field algebra, independently of how its positive quasifree covariance is subsequently constructed.

\subsection{Positive two-point functions on the unstable band}
\label{subsec:positive-unstable}

Now, we investigate how the unstable state can be incorporated into the positive two-point function \eqref{eq:W-symmetric-decomposition}, while keeping the stable contribution~\eqref{eq:stable-wightman}. For this purpose, we restrict attention to states invariant under spatial translations parallel to the boundary, with real and even Cauchy covariances in tangential momentum space. The symmetric part of the complete two-point function can then be written as
\begin{equation}
 S(x,x')=\frac12\left[W_{\mathrm{st}}(x,x')+
 W_{\mathrm{st}}(x',x)\right]+S_{\mathrm{un}}(x,x'),
\end{equation}
where
\begin{equation}
    S_{\mathrm{un}}(x,x'):=\psi_b(z)\psi_b(z')
 \int_{k<\beta}\frac{\dd^{2}\kb}{(2\pi)^{2}}
 \ee^{\ii\kb\cdot(\xb-\xpb)}s_{k}(t,t').
 \label{eq:unstable-symmetric-kernel}
\end{equation}
Here, $s_k$ is real, even under $\kb\mapsto-\kb$, and symmetric under the exchange $t\leftrightarrow t'$.

For the reference surface $\Sigma_0\times \Sigma_0$, we parametrize the Cauchy data of $s_{k}$ by
\begin{equation}
\begin{aligned}
 A_{k}&:=\left.s_{k}(t,t')\right|_{t=t'=0},\\
 B_{k}&:=\left.\partial_t s_{k}(t,t')\right|_{t=t'=0}
     =\left.\partial_{t'}s_{k}(t,t')\right|_{t=t'=0},\\
 C_{k}&:=\left.\partial_t\partial_{t'}s_{k}(t,t')\right|_{t=t'=0}.
\end{aligned}
\label{eq:covariances}
\end{equation}
The antisymmetric part is not an additional choice, instead, it is given by Eq.~\eqref{eq:full-E}, with the reduced kernel
\begin{equation}
 e_k(t,t'):=-\frac{\sinh[\gamma_k(t-t')]}{\gamma_k}.
 \label{eq:reduced-unstable-propagator}
\end{equation}
Accordingly, the complete reduced two-point function must have the form
\begin{equation}
 w_{k}(t,t')=s_{k}(t,t')+\frac{\ii}{2}e_k(t,t'),
 \label{eq:reduced-W-decomposition}
\end{equation}
with the initial data on the antisymmetric part being
\begin{equation}
\begin{aligned}
 \left.e_k(t,t')\right|_{t=t'=0}&=0,
 &\left.\partial_{t'}e_k(t,t')\right|_{t=t'=0}&=1,\\
 \left.\partial_t e_k(t,t')\right|_{t=t'=0}&=-1,
 &\left.\partial_t\partial_{t'}e_k(t,t')\right|_{t=t'=0}&=0.
\end{aligned}
\label{eq:reduced-E-Cauchy-data}
\end{equation}

Consequently, the Cauchy covariance of $w_{k}$ is the Hermitian matrix
\begin{equation}
 \Gamma_{k}:=\left.
 \begin{pmatrix}
 w_{k}&\partial_{t'}w_{k}\\
 \partial_t w_{k}&\partial_t\partial_{t'}w_{k}
 \end{pmatrix}\right|_{t=t'=0}=\begin{pmatrix}
 A_{k}&B_{k}+\ii/2\\
 B_{k}-\ii/2&C_{k}
 \end{pmatrix}.
\label{eq:initial-covariance-matrix}
\end{equation}
Notice that the imaginary entries $\pm\ii/2$ are fixed by the CCR, while the real coefficients $A_k$, $B_k$, and $C_k$ encode the freedom in choosing the state.

We now impose positivity on this covariance. After tangential Fourier transformation, positivity requires $\Gamma_k$ to be positive semidefinite for almost every momentum in the unstable band. Since $\Gamma_k$ is a Hermitian $2\times2$ matrix, this condition is equivalent to the non-negativity of its principal minors, giving
\begin{equation}
 A_{k}\geq0,\qquad C_{k}\geq0,\qquad
 A_kC_{k}-B_{k}^2\geq\frac14.
 \label{eq:uncertainty}
\end{equation}
These inequalities are precisely the Robertson--Schrödinger uncertainty relations~\cite{Robertson1929}, applied mode by mode to the unstable sector. They are also the $2\times2$ realization of the general uncertainty condition for quasifree states~\cite{KayWald1991}.

Although Eq.~\eqref{eq:uncertainty} constrains the admissible Cauchy covariances, we must still establish that their evolution defines a positive two-point function after integration over the unstable momentum band. To this end, we impose the sufficient integrability condition
\begin{equation}
 A_{k},B_{k},C_{k}\in
 L^1\!\left(\{\kb:k<\beta\},\dd^{2}\kb\right).
 \label{eq:unstable-admissibility}
\end{equation}
This assumption includes the behavior arbitrarily close to the threshold $k=\beta$ and will also allow us to establish the smoothness of the resulting two-point function.

Since each unstable momentum satisfies the temporal equation $\ddot \chi_k-\gamma_k^2 \chi_k=0$, its Cauchy data evolve according to the fundamental matrix
\begin{equation}
T_k(t):=
\begin{pmatrix}
\cosh(\gamma_k t)&\ \sinh(\gamma_k t)/\gamma_k\\
\gamma_k\sinh(\gamma_k t)&\cosh(\gamma_k t)
\end{pmatrix},
\label{eq:unstable-fundamental-matrix} 
\end{equation}
which satisfies $T_k(0)=\1$. In particular, defining
\begin{equation}
\mathbf h_k(t):=
\begin{pmatrix}
\cosh(\gamma_k t)\\
\sinh(\gamma_k t)/\gamma_k
\end{pmatrix},
\end{equation}
we can express the evolution of the reduced two-point function as
\begin{equation}
w_k(t,t')=\mathbf h_k(t)^T\Gamma_k\mathbf h_k(t').
\label{eq:unstable-Cauchy-evolution}
\end{equation}
Here, the ratios involving $\gamma_k$ are understood through their continuous limits as $\gamma_k\to0$, in particular $\sinh(\gamma_k t)/\gamma_k\to t$, ensuring that the evolution is regular at the threshold.

Expanding Eq.~\eqref{eq:unstable-Cauchy-evolution}, we find that its real part gives the symmetric covariance
\begin{equation}
\begin{aligned}
&s_k(t,t')={}
\frac12\left(A_k+\frac{C_k}{\gamma_k^2}\right)
\cosh[\gamma_k(t+t')]\\
&\,\,{}+\frac{B_k}{\gamma_k}\sinh[\gamma_k(t+t')]+\frac12\left(A_k-\frac{C_k}{\gamma_k^2}\right)
\cosh[\gamma_k(t-t')].
\end{aligned}
\label{eq:symmetric-kernel-unstable}
\end{equation}
Thus, the same evolution propagates the freely specified symmetric covariance while preserving the antisymmetric contribution fixed by the CCR.

The construction above provides a positive covariance for each unstable momentum. To obtain a quantum state, however, we must show that their momentum integral defines a positive-type distribution with the required regularity and boundary properties. This is established by the following proposition.

\begin{proposition}[Unstable completion]
\label{prop}
Let $\beta>0$, and let $A_k$, $B_k$, and $C_k$ be real, even, measurable covariance data satisfying Eqs.~\eqref{eq:uncertainty} and \eqref{eq:unstable-admissibility}. Then
\begin{equation}
W_{\mathrm{un}}(x,x'):=\psi_b(z)\psi_b(z')\int_{k<\beta}\frac{\dd^{2}\kb}{(2\pi)^{2}}\ee^{\ii\kb\cdot(\xb-\xpb)}w_k(t,t')
\label{eq:unstable-two-point}
\end{equation}
is a smooth positive-type bisolution satisfying the Robin boundary condition in both arguments. Its antisymmetric part is $\ii E_{\mathrm{un}}$, where $E_{\mathrm{un}}$ denotes the $k<\beta$ contribution to Eq.~\eqref{eq:full-E}. 
\end{proposition}

\noindent\textit{Proof.}
We first establish positivity. Under the evolution~\eqref{eq:unstable-Cauchy-evolution}, the equal-time Cauchy covariance transforms according to
\begin{equation}
\Gamma_k(t)=T_k(t)\Gamma_kT_k(t)^T,
\label{eq}
\end{equation}
which preserves positive semidefiniteness. Moreover, using $\cosh^2(\gamma_k t)-\sinh^2(\gamma_k t)=1$, we find $\det T_k(t)=1$, ensuring that the antisymmetric Cauchy form is preserved. 

However, positivity on individual surfaces is not sufficient by itself to establish positivity of the complete two-point kernel at different times. To this end, for a complex test function $f$, define
\begin{equation}
\mathbf F_f(\kb):= \int\dd^4 x\,\ee^{-\ii\kb\cdot\xb}\psi_b(z)\mathbf h_k(t)f(x).
\end{equation}
Then
\begin{equation}
W_{\mathrm{un}}(f^*,f)=\int_{k<\beta}\frac{\dd^{2}\kb}{(2\pi)^{2}}\mathbf F_f(\kb)^\dagger\Gamma_k\mathbf F_f(\kb)\geq0.
\label{eq}
\end{equation}
The inequality follows from the positive semidefiniteness of $\Gamma_k$, while the integral is finite by Eq.~\eqref{eq:unstable-admissibility} and the boundedness of the hyperbolic evolution functions on compact time intervals. Therefore, within the covariance ansatz considered here, Eq.~\eqref{eq:uncertainty} is both necessary and sufficient for positivity.

Although the two-point function is initially defined on $M^\circ\times M^\circ$, the unstable contribution extends smoothly to $M\times M$. Let $K\subset M\times M$ be compact, and let $D$ denote any finite-order spacetime derivative in the two arguments. Since $0\leq\gamma_k\leq\beta$, the hyperbolic functions $\cosh(\gamma_k t)$ and $\sinh(\gamma_k t)/\gamma_k$, together with all their time derivatives, are uniformly bounded on compact time intervals, including as $\gamma_k\to0$. Furthermore, tangential derivatives produce powers of $\kb$, which remain bounded on the unstable momentum ball, while normal derivatives of $\psi_b(z)\psi_b(z')$ introduce only powers of $\beta$. Therefore, there exists a constant $c_{K,D}$, independent of $\kb$, such that
\begin{equation}
\begin{aligned}
&\left|D\left[
\psi_b(z)\psi_b(z')
\ee^{\ii\kb\cdot(\xb-\xpb)}w_k(t,t')
\right]\right|\\
&\hspace{10mm}\leq
c_{K,D}\bigl(|A_k|+|B_k|+|C_k|+1\bigr).
\end{aligned}
\label{eq}
\end{equation}
The right-hand side is integrable on the unstable ball. Thus, every derivative can be passed under the integral, yielding
\begin{equation}
W_{\mathrm{un}}\in C^\infty(M\times M).
\end{equation}
Notice that this result is local in time and does not require the kernel or its derivatives to remain bounded as $|t|\to\infty$.

Finally, the hyperbolic time-dependent functions solve the unstable temporal equation, while $\psi_b$ satisfies the Robin boundary condition. As the antisymmetric part of Eq.~\eqref{eq:unstable-Cauchy-evolution} reproduces Eq.~\eqref{eq:reduced-unstable-propagator}, these properties establish that $W_{\mathrm{un}}$ is a bisolution satisfying the required boundary condition and possessing the antisymmetric part $\ii E_{\mathrm{un}}$.
\hfill$\square$

Having established the existence of a positive unstable contribution, we can combine it with the stable two-point function to obtain
\begin{equation}
W(x,x')=W_{\mathrm{st}}(x,x')+W_{\mathrm{un}}(x,x').
\label{eq:full-two-point}
\end{equation}
Both contributions are of positive type and satisfy the field equation and the Robin boundary condition. Moreover, their antisymmetric parts reconstruct the complete causal propagator obtained in Sec.~\ref{subsec:continuum}, ensuring compatibility with the CCR. Therefore, Eqs.~\eqref{eq:uncertainty} and \eqref{eq:unstable-admissibility} provide a family of positive quasifree states on the complete field algebra, including the unstable spectral sector. 

\subsection{Obstruction to stationarity}
\label{subsec:stationarity}

The stable contribution $W_{\mathrm{st}}$ in Eq.~\eqref{eq:stable-wightman} is stationary, so the complete two-point function in Eq.~\eqref{eq:full-two-point} can be stationary only if its unstable contribution is also invariant under time translations. To investigate this condition, notice that $w(t,t')$ contains terms depending on both $t+t'$ and $t-t'$, whereas the antisymmetric contribution~\eqref{eq:reduced-unstable-propagator} depends only on their difference. Accordingly, stationarity requires
\begin{equation}
s_k(t+\tau,t'+\tau)=s_k(t,t'),\qquad \text{for all }\tau,
\label{eq}
\end{equation}
for almost every momentum in the unstable band.

The above requirement is satisfied if the terms depending on $t+t'$ vanish on Eq.~\eqref{eq:symmetric-kernel-unstable}. Hence, this means
\begin{equation}
B_k=0,\qquad C_k=-\gamma_k^2 A_k.
\label{eq:stationarity-condition}
\end{equation}
Considering the uncertainty relation, Eq.~\eqref{eq:stationarity-condition} yields
\begin{equation}
A_kC_k-B_k^2=-\gamma_k^2 A_k^2\leq0,
\end{equation}
in contradiction with positivity in Eq.~\eqref{eq:uncertainty}. Therefore, no positive quasifree state in the covariance class constructed in Sec.~\ref{subsec:positive-unstable} can be stationary while satisfying the complete CCR. Although the stable two-point function $W_{\mathrm{st}}$ is stationary and of positive type, retaining it alone would omit the unstable projector~\eqref{eq:missing-unstable-projector} and therefore violate the CCR of the complete field.

Despite this temporal growth, the energy associated with the unstable sector remains conserved. Using the symmetric covariance to evaluate the quadratic Hamiltonian, we obtain the unstable contribution to the energy per unit tangential volume:
\begin{equation}
\begin{aligned}
 \mathcal E[W](t)
 &:=\frac12\int_{k<\beta}\frac{\dd^{2}\kb}{(2\pi)^{2}}
 \left[\partial_t\partial_{t'}s_{k}-\gamma_k^2s_{k}\right]_{t'=t}\\
 &=\frac12\int_{k<\beta}\frac{\dd^{2}\kb}{(2\pi)^{2}}
 \left(C_{k}-\gamma_k^2A_{k}\right).
\end{aligned}
 \label{eq:energy-covariance}
\end{equation}
The second equality follows directly from Eq.~\eqref{eq:covariances} and shows that the energy is independent of time. Moreover, since $\gamma_k\leq\beta$, its finiteness follows from the integrability condition \eqref{eq:unstable-admissibility}.

Nevertheless, conservation does not imply that this energy is bounded below. Considering the admissible covariance data $A_k=a>0$, $B_k=0$, and $C_k=1/(4a)$, for $a=\,$constant, it saturate the uncertainty relation~\eqref{eq:uncertainty} and satisfy the integrability condition~\eqref{eq:unstable-admissibility}. However, as $a\to\infty$, the energy in Eq.~\eqref{eq:energy-covariance} goes to $-\infty$. Therefore, an exponentially growing fluctuation on the unstable sector is compatible with the conservation of its indefinite energy.

\section{Complete two-point function and the Hadamard property}
\label{sec:hadamard}

Proposition~\ref{prop} establishes that the unstable contribution to the two-point function is smooth. Therefore, although its inclusion is necessary to recover the complete CCR, it does not introduce additional singularities into the two-point function. It remains, however, to investigate the singular structure due to the stable contribution to identify a common, state-independent singularity representative within the family of states constructed above. Rather than constructing a new Hadamard parametrix, our purpose is to identify the singularities of the complete two-point function and compare them with the Robin--Hadamard structure established in Ref.~\cite{CosteriEtAl2026}.

To understand the geometrical origin of these singularities, let us recall how the timelike boundary modifies their interpretation. In the absence of a boundary, the local Hadamard structure is characterized by Synge's world function $\sigma(x,x')$, defined as one-half of the squared geodesic interval between two points within a convex normal neighborhood. In flat space, this geodesic is simply a straight line, so that
\begin{equation}
\sigma(x,x')=\frac{1}{2}\left[-\tau^2+r^2+(z-z')^2\right],
\label{eq:synge}
\end{equation}
for $\tau:=t-t'$ and $r:=|\xb-\xpb|$. In particular, $\sigma(x,x')=0$ characterizes pairs of distinct points connected by a direct null geodesic within the neighborhood.

In the presence of the timelike boundary, however, singularities may also propagate along null rays that reach the boundary and are reflected back into the spacetime. These trajectories can be described by introducing the mirror image of $x'$ across the boundary: $\bar{x}'=(t',\xpb,-z')$. The straight line connecting $x$ to $\bar{x}'$ is the unfolded representation of a trajectory reflected at $\partial M$. Accordingly, we introduce the reflected world function
\begin{equation}
\bar\sigma(x,x')=\frac12\left[-\tau^2+r^2+(z+z')^2\right].
\label{eq:reflected}
\end{equation}
Thus, while $\sigma=0$ describes direct null propagation, $\bar\sigma=0$ characterizes the corresponding boundary-reflected null trajectories. Notice that $\bar\sigma$ is not Synge's world function for a geodesic connecting $x$ and $x'$ in $M^\circ$, but rather the world function of the ambient Minkowski geometry evaluated at the reflected point.

In what follows, we simplify the notation by considering the normal separations $s:=z-z'$ and $a:=z+z'$, together with $\zeta:=\sqrt{r^2-\tau^2}$. Since the two-point function is obtained from positive-frequency mode integrals in the stable sector, we also introduce the Wightman prescription $\tau_\epsilon:=\tau-\ii\epsilon$, with $\epsilon>0$, and define
\begin{equation}
\zeta_\epsilon
:=\sqrt{r^2-\tau_\epsilon^2},
\end{equation}
where the square root is chosen with $\operatorname{Re}\zeta_\epsilon>0$. With this prescription, Eqs.~\eqref{eq:synge} and \eqref{eq:reflected} take the compact form
\begin{equation}
2\sigma_\epsilon=\zeta_\epsilon^2+s^2,
\qquad
2\bar\sigma_\epsilon=\zeta_\epsilon^2+a^2,
\label{eq:regularized-world-function}
\end{equation}
where $\sigma_\epsilon=\sigma+\ii\epsilon\tau+\epsilon^2/2$, with an analogous expression for $\bar\sigma_\epsilon$. 


\subsection{The continuum contribution}
\label{subsec:hadamard-continuum}

We begin by evaluating $W_c(x,x')$ in position space. Using the normalized modes~\eqref{eq:continuum-modes}, and considering an auxiliary real Euclidean time separation $T>0$, Eq.~\eqref{eq:continuous-wightman} can be written as
\begin{equation}
 W_c^{\mathrm E}(x,x')
 =\int_0^\infty\dd q\,
 \mathcal F_T(q,r)\rho_q(z)\rho_q(z'),
 \label{eq:hadamard-continuum-euclidean}
\end{equation}
where $\mathcal F_T$ denotes the Fourier integral over the momenta parallel to the boundary
\begin{equation}
\begin{aligned}
 \mathcal F_T(q,r)
 &:=\int_{\R^2}\frac{\dd^2\kb}{(2\pi)^2}
 \frac{\ee^{\ii\kb\cdot(\xb-\xpb)}
 \ee^{-T\sqrt{k^2+q^2}}}{2\sqrt{k^2+q^2}}\\
 &=\frac{1}{4\pi}\int_0^\infty\dd k\,
 \frac{kJ_0(kr)}{\sqrt{k^2+q^2}}
 \ee^{-T\sqrt{k^2+q^2}}.
\end{aligned}
\label{eq:hadamard-tangential-fourier}
\end{equation}
Here, $J_0$ is the Bessel function of the first kind, arising from the angular integration in tangential momentum space.

To evaluate the remaining radial integral, we use the Schwinger-parameter representation
\begin{equation}
 \frac{\ee^{-T\sqrt{k^2+q^2}}}{2\sqrt{k^2+q^2}}
 =\frac{1}{2\sqrt\pi}\int_0^\infty
 \frac{\dd u}{\sqrt u}\,
 \exp\!\left[-u(k^2+q^2)-\frac{T^2}{4u}\right],
\label{eq:hadamard-schwinger}
\end{equation}
which is the $\nu=1/2$ case of Eq.~3.471.9 in Ref.~\cite{gradshteyn2014table}, together with the elementary expression for $K_{1/2}$. For the Bessel function contribution we can use the Gaussian--Bessel identity (see Eq.~10.22.51 in Ref.~\cite{DLMF})
\begin{equation}
 \int_0^\infty\dd k\,kJ_0(kr)\ee^{-uk^2}
 =\frac{1}{2u}\ee^{-r^2/(4u)},
 \qquad u>0.
\label{eq:hadamard-gaussian-bessel}
\end{equation}

Substituting Eqs.~\eqref{eq:hadamard-schwinger} and \eqref{eq:hadamard-gaussian-bessel} into Eq.~\eqref{eq:hadamard-tangential-fourier}, the remaining Schwinger-parameter integral gives
\begin{equation}
 \mathcal F_T(q,r)=\frac{\ee^{-qR}}{4\pi R},
 \qquad R:=\sqrt{r^2+T^2}.
 \label{eq:hadamard-F-result}
\end{equation}
Thus, the tangential Fourier integration reduces the problem to a single integral over the normal spectral parameter $q$.

The product of eigenfunctions~\eqref{eq:rhoq} allows the expansion
\begin{equation}
\begin{aligned}
 \rho_q(z)\rho_q(z')
 =\frac{1}{\pi}\biggl[
 &\cos(qs)+\cos(qa)\\
 &-2\beta
 \frac{\beta\cos(qa)+q\sin(qa)}{q^2+\beta^2}
 \biggr],
\end{aligned}
\label{eq:hadamard-rho-product}
\end{equation}
where $s=z-z'$ and $a=z+z'$, as introduced previously. Notice that the first two terms already distinguish the direct and reflected normal separations, while the last term contains the explicit dependence on the Robin parameter.

Indeed, using the elementary Laplace transform
\begin{equation}
 \int_0^\infty\dd q\,
 \ee^{-qR}\cos(qb)
 =\frac{R}{R^2+b^2},
 \qquad R>0,
\end{equation}
we see that the first two contributions reproduce the direct and reflected massless kernels. The Robin-dependent term can be treated similarly by introducing the integral representation
\begin{equation}
 \frac{\beta\cos(qa)+q\sin(qa)}{q^2+\beta^2}
 =\int_0^\infty\dd v\,
 \ee^{-\beta v}\cos[q(a-v)].
\label{eq:hadamard-robin-convolution}
\end{equation}
This allows us to see the Robin contribution as a weighted superposition of the same cosine transform, now evaluated at the shifted normal separation $a-v$.

Accordingly, the $q$ integration can be performed using the same identity, and Eq.~\eqref{eq:hadamard-continuum-euclidean} becomes
\begin{equation}
 W_c^{\mathrm E}
 =\frac{1}{4\pi^2(R^2+s^2)}
 +\frac{1}{4\pi^2(R^2+a^2)}
 -\frac{\beta}{2\pi^2}I_\beta(R,a),
\label{eq:hadamard-continuum-position}
\end{equation}
where 
\begin{equation}
 I_\beta(R,a):=
 \int_0^\infty\frac{\ee^{-\beta v}\,\dd v}
 {R^2+(a-v)^2}.
\label{eq:hadamard-I-beta}
\end{equation}
As $R>0$ and $\beta>0$, this integral is absolutely convergent. Notice also that the factor $R$ obtained from the Laplace transform cancels the denominator $R$ in Eq.~\eqref{eq:hadamard-F-result}.

Equation~\eqref{eq:hadamard-continuum-position} already makes the geometrical content of the continuum contribution explicit. After the Lorentzian continuation $T\to T_\epsilon=\ii\tau_\epsilon$, the first two denominators become $2\sigma_\epsilon$ and $2\bar\sigma_\epsilon$, respectively. The remaining term encodes the additional dependence on the Robin boundary parameter. Decomposing the denominators of $I_{\beta}$ we get
\begin{equation}
 \frac{1}{R^2+(a-v)^2}
 =\frac{1}{2\ii R}
 \left[
 \frac{1}{v-a-\ii R}
 -\frac{1}{v-a+\ii R}
 \right],
\label{eq:hadamard-I-partial-fractions}
\end{equation}
which allows us to evaluate the integrals using the identity (Eq.~6.7.1 of Ref.~\cite{DLMF})
\begin{equation}
 \int_0^\infty\frac{\ee^{-\beta v}}{v+c}\,\dd v
 =\ee^{\beta c}E_1(\beta c),
\label{eq:hadamard-E1-transform}
\end{equation}
initially for $c>0$ and then by analytic continuation to $c\in\mathbb C\setminus(-\infty,0]$. Here, $E_1$ is the exponential integral. Accordingly, we obtain
\begin{equation}
\begin{aligned}
 I_\beta(R,a)=\frac{\ee^{-\beta a}}{2\ii R}\Bigl[
 &\ee^{-\ii\beta R}E_1(-\beta a-\ii\beta R)\\
 &-\ee^{\ii\beta R}E_1(-\beta a+\ii\beta R)
 \Bigr].
\end{aligned}
\label{eq:hadamard-I-E1}
\end{equation}

At this stage, the branch structure of the exponential integral becomes important. For real $R>0$ and $a>0$, the two arguments of $E_1$ lie on opposite sides of its negative-real branch cut. Using the principal branches and the corresponding analytic continuation gives us
\begin{equation}
\begin{aligned}
 E_1(-\beta a-\ii\beta R)
 &=-\operatorname{Ei}(\beta a+\ii\beta R)+\ii\pi,\\
 E_1(-\beta a+\ii\beta R)
 &=-\operatorname{Ei}(\beta a-\ii\beta R)-\ii\pi.
\end{aligned}
\label{eq:hadamard-E1-Ei-branches}
\end{equation}
Finally, we have 
\begin{equation}
\begin{aligned}
 I_\beta(R,a)={}&
 \frac{\pi\ee^{-\beta a}}{R}\cos(\beta R)\\
 &-\frac{\ee^{-\beta a}}{2\ii R}\Bigl[
 \ee^{-\ii\beta R}\operatorname{Ei}(\beta a+\ii\beta R)\\
 &\hspace{20mm}
 -\ee^{\ii\beta R}
 \operatorname{Ei}(\beta a-\ii\beta R)
 \Bigr].
\end{aligned}
\label{eq:hadamard-I-Ei}
\end{equation}

Having found a closed form on the oscillatory sector, we now turn to the bound state.

\subsection{The bound-state contribution}
\label{subsec:hadamard-bound}

Our purpose is to evaluate the stable and unstable bound contributions by considering the complete two-point function of the former and only the antisymmetric part of the latter, leaving the freedom of a state choice in $S_{\mathrm{un}}$ explicit.

The stable contribution admits the expansion
\begin{equation}
W_b
=\frac{\beta\ee^{-\beta a}}{2\pi}
\int_0^\infty\dd\Omega
J_0\left(r\sqrt{\Omega^2+\beta^2}\right)
\ee^{-\ii\Omega\tau_\epsilon},
\label{eq:hadamard-stable-bound-radial}
\end{equation}
where we have integrated the angular dependence parallel to $\xb-\xpb$ and considering the variable changes $k\dd k/\Omega=\dd\Omega$. Concerning the antisymmetric part of the two-point correlation function associated to the unstable modes, the analogous integration yields
\begin{equation}
\begin{aligned}
 \frac{\ii}{2}E_{\mathrm{un}}
 =-\frac{\ii\beta\ee^{-\beta a}}{2\pi}
 \int_0^\beta\dd\gamma\,
 J_0\!\left(r\sqrt{\beta^2-\gamma^2}\right)
 \sinh(\gamma\tau).
\end{aligned}
\label{eq:hadamard-unstable-radial}
\end{equation}
In this latter case, the momentum band is bounded and no regulator $i\epsilon$ is required, in accordance with our proposition \ref{prop}. Nevertheless, its dependence on $\tau$ admits an analytic continuation, which will be useful when combining it with Eq.~\eqref{eq:hadamard-stable-bound-radial}.

Now, return to the tangential Fourier transform~\eqref{eq:hadamard-tangential-fourier}--\eqref{eq:hadamard-F-result}. Observe that the square root $\sqrt{k^2+q^2}$ admits two boundary values, corresponding to $q=\pm\ii\beta$, obtained by approaching the negative real axis of $q^2$ from opposite sides of the branch cut. Averaging, we find
\begin{equation}
\frac{\mathcal F_T(+\ii\beta,r)
+\mathcal F_T(-\ii\beta,r)}{2}
=\frac{\cos(\beta R)}{4\pi R}.
\label{eq:media}
\end{equation}
For $k>\beta$, both continuations of $\sqrt{k^2+q^2}$ give the positive frequency $\Omega_k=\sqrt{k^2-\beta^2}$, recovering the Euclidean stable bound-mode factor $\exp(-T\Omega_k)/(2\Omega_k)$. For $k<\beta$, however, the square roots approach $\pm\ii\gamma_k$, and their average gives
\begin{equation}
\frac12\left[
\frac{\ee^{-\ii\gamma_kT}}{2\ii\gamma_k}
+\frac{\ee^{\ii\gamma_kT}}{-2\ii\gamma_k}
\right]
=-\frac{\sin(\gamma_kT)}{2\gamma_k}.
\label{eq:media2}
\end{equation}
Although the individual terms have integrable singularities at $k=\beta$, their average remains finite as $\gamma_k\to0$.

To recover the Lorentzian expression, we analytically continue Eq.~\eqref{eq:media2} to
$T=\ii\tau_\epsilon$ obtaining $\ii e_k/2$, where $e_k$ is defined in Eq.~\eqref{eq:reduced-unstable-propagator}. Therefore, the two momentum regions of the averaged Fourier transform can be identified explicitly by writing
\begin{equation}
\begin{aligned}
&\psi_b(z)\psi_b(z')
\frac{\mathcal F_T(+\ii\beta,r)+\mathcal F_T(-\ii\beta,r)}{2}\\
&\quad=\frac{\beta\ee^{-\beta a}}{2\pi}
\biggl[
\int_0^\infty\dd\Omega\,
J_0\left(r\sqrt{\Omega^2+\beta^2}\right)
\ee^{-T\Omega}\\
&\qquad\qquad-
\int_0^\beta\dd\gamma\,
J_0\left(r\sqrt{\beta^2-\gamma^2}\right)
\sin(\gamma T)
\biggr].
\end{aligned}
\end{equation}
The first integral corresponds to the stable bound contribution, while the second reproduces the antisymmetric unstable contribution fixed by the CCR. Hence, continuing the combined expression to $T=\ii\tau_\epsilon$ and using Eq.~\eqref{eq:media2} we obtain
\begin{equation}
 W_b+\frac{\ii}{2}E_{\mathrm{un}}
 =\frac{\beta\ee^{-\beta a}}{2\pi\zeta_\epsilon}
   \cos(\beta\zeta_\epsilon),
\label{eq:hadamard-bound-combination}
\end{equation}
(remember the limit $\epsilon\downarrow0$ is smooth for the unstable kernel).

Combining \eqref{eq:hadamard-bound-combination} with \eqref{eq:hadamard-I-Ei}, the cosine contributions cancel, yielding the complete two-point function in position space:
\begin{equation}
W(x,x')=\lim_{\epsilon\downarrow0}\biggl[\frac{1}{8\pi^2\sigma_\epsilon}
+\frac{1}{8\pi^2\bar\sigma_\epsilon}+\mathcal R_\beta(\zeta_\epsilon,a)
\biggr]+S_{\mathrm{un}}(x,x'),
\label{eq:final-two-point}
\end{equation}
with
\begin{equation}
\begin{aligned}
 \mathcal R_\beta(\zeta,a)
 :=\frac{\beta\ee^{-\beta a}}{4\pi^2\ii\zeta}\Bigl[&
 \ee^{-\ii\beta\zeta}
 \operatorname{Ei}(\beta a+\ii\beta\zeta)\\
 &-\ee^{\ii\beta\zeta}
 \operatorname{Ei}(\beta a-\ii\beta\zeta)
 \Bigr].
\end{aligned}
\label{eq:hadamard-Robin-kernel}
\end{equation}
The limit is understood in the distributional sense, with the branches fixed by the preceding analytic continuation. The remaining term $S_{\mathrm{un}}$ is the smooth symmetric covariance defined in Eq.~\eqref{eq:unstable-symmetric-kernel}, whose explicit form depends on the choice of $A_k$, $B_k$, and $C_k$. Thus, all states in the constructed family share the same state-independent singularity structure.

\subsection{Singularity structure and state-independent subtraction}
\label{subsec:robin-hadamard-parametrix}

Having obtained the complete two-point function in position space, we can now examine its singularity structure and compare it with the Robin--Hadamard form discussed in Ref.~\cite{CosteriEtAl2026}. From Sec.~\ref{subsec:positive-unstable}, the unstable contribution is smooth and do not affect the singular structure at the coincident limit $x'\to x$. This motivates us to identify the subtraction kernel of Eq.~\eqref{eq:final-two-point} as
\begin{equation}
 H_\beta(x,x'):=\lim_{\epsilon\to0^+}\left[
 \frac{1}{8\pi^2\sigma_\epsilon}
 +\frac{1}{8\pi^2\bar\sigma_\epsilon}
 +\mathcal R_\beta(\zeta_\epsilon,a)\right].
\label{eq:hadamard-spectral-representative}
\end{equation}
Rather than constructing this kernel through Hadamard recursion relations, we have obtained it directly from the spectral representation of the complete two-point function. We now verify that it exhibits the expected singularity structure and satisfies the dynamical and boundary conditions of the Robin problem.

The first two terms in Eq.~\eqref{eq:hadamard-spectral-representative} reproduce the direct and reflected massless singularities, respectively, associated with $\sigma=0$ and $\bar\sigma=0$. The remaining contribution $\mathcal R_\beta$, however, contains additional terms whose singular behavior is not immediately apparent from its representation in terms of exponential integrals. To identify them, we recall the expansion (Eqs.~6.6.1--6.6.2 of Ref.~\cite{DLMF})
\begin{equation}
 \operatorname{Ei}(w)
 =\gamma_{\mathrm E}+\operatorname{Log}w
 +\sum_{j=1}^\infty\frac{w^j}{j\,j!}.
\label{eq:hadamard-Ei-series}
\end{equation}
Since the series is entire, the nonanalytic contributions arise from the logarithms when $a+\ii\zeta$ or $a-\ii\zeta$ vanishes. Notice that
\begin{equation}
 (a+\ii\zeta)(a-\ii\zeta)
 =a^2+\zeta^2=2\bar\sigma,
\end{equation}
so these singularities occur precisely on the reflected null relation $\bar\sigma=0$.

Considering a neighborhood of a reflected null-related pair in the spacetime interior, introduce
\begin{equation}
 d(\bar\sigma,a):=\sqrt{a^2-2\bar\sigma},
 \qquad d(0,a)=a.
\label{eq:hadamard-reflected-local-root}
\end{equation}
Using Eq.~\eqref{eq:hadamard-Robin-kernel}, we can isolate the logarithmic contribution in the form
\begin{equation}
 \mathcal R_\beta(\zeta_\epsilon,a)
 =\mathcal V_\beta(\bar\sigma_\epsilon,a)
 \operatorname{Log}\!\left(
 \frac{2\bar\sigma_\epsilon}{\ell^2}\right)
 +\mathcal Q_{\beta,\epsilon},
\label{eq:hadamard-reflected-log-form}
\end{equation}
where $\ell>0$ is an arbitrary reference length, $\mathcal Q_{\beta,\epsilon}$ admits a smooth boundary value in the neighborhood, and
\begin{equation}
 \mathcal V_\beta(\bar\sigma,a)
 =-\frac{\beta}{4\pi^2d(\bar\sigma,a)}
 \exp\!\left[-\beta\bigl(a-d(\bar\sigma,a)\bigr)\right].
\label{eq:hadamard-reflected-log-coefficient}
\end{equation}
In particular, on the reflected null surface we find
\begin{equation}
 \mathcal V_\beta(0,a)=-\frac{\beta}{4\pi^2a}.
\end{equation}
Thus, the Robin parameter introduces a logarithmic singularity associated with the reflected null rays, in addition to the usual reflected pole. The dependence of $\mathcal V_\beta$ on $\bar\sigma$ generates an expansion in powers of $\bar\sigma$ multiplying the logarithm. Although terms of the form $\bar\sigma^n\operatorname{Log}\bar\sigma$ vanish as $\bar\sigma\to0$ for $n\geq1$, their derivatives need not remain regular. Such contributions may therefore become relevant when considering renormalized local observables involving derivatives of the two-point function. In the present analysis, however, retaining the complete coefficient $\mathcal V_\beta$ is sufficient to isolate the logarithmic contribution with a smooth remainder. This expansion is valid locally for $a>0$ and is not uniform as $a\to0$, where the direct and reflected singularities meet.

On the other hand, $\mathcal R_\beta$ introduces no additional singularity at $\zeta=0$ for $a>0$. Indeed, the numerator of Eq.~\eqref{eq:hadamard-Robin-kernel} is odd and analytic in $\zeta$ near the origin, so division by $\zeta$ yields a regular function of $\zeta^2$. Consequently, near any coincidence point in the spacetime interior, the singularity is governed by the standard direct Hadamard term $1/(8\pi^2\sigma_\epsilon)$, while the reflected contributions remain smooth in a sufficiently small convex normal neighborhood. Away from coincidence, the additional singularities arise along reflected null trajectories, consisting of the reflected pole and the Robin-dependent logarithmic terms identified above. These contributions exhibit the direct and reflected singularity structure expected from the Robin--Hadamard framework of Ref.~\cite{CosteriEtAl2026}, with the Wightman prescription fixing their frequency orientation.

Besides the singularity structure, the subtraction kernel must also be compatible with the field dynamics and the CCR. From its spectral construction, we have
\begin{equation}
 H_\beta=W_c+W_b+\frac{\ii}{2}E_{\mathrm{un}}.
\label{eq:hadamard-H-spectral-identity}
\end{equation}
Since each contribution is a bisolution of the wave equation, it follows that
\begin{equation}
 \Box_xH_\beta=\Box_{x'}H_\beta=0.
\label{eq:hadamard-H-bisolution}
\end{equation}
Moreover, their antisymmetric parts reproduce the complete causal propagator, i.e.,
\begin{equation}
 H_\beta(x,x')-H_\beta(x',x)
 =\ii E(x,x').
\label{eq:hadamard-H-CCR}
\end{equation}
Thus, the subtraction kernel preserves the commutation relations of the complete field, including the unstable contribution required by spectral completeness.

The Robin boundary condition also follows from the spectral identity~\eqref{eq:hadamard-H-spectral-identity}, but it can be verified directly from the position-space representation. Indeed, integrating Eq.~\eqref{eq:hadamard-I-beta} by parts gives
\begin{equation}
 (\partial_a+\beta)I_\beta(R,a)
 =\frac{1}{R^2+a^2},
\label{eq:hadamard-I-robin-identity}
\end{equation}
which, after the cancellation of the cosine terms and analytic continuation, implies
\begin{equation}
 (\partial_a+\beta)\mathcal R_\beta(\zeta_\epsilon,a)
 =-\frac{\beta}{2\pi^2(\zeta_\epsilon^2+a^2)}.
\label{eq:hadamard-R-robin-identity}
\end{equation}
At $z=0$, the normal derivatives of the direct and reflected massless contributions cancel, while the terms proportional to $\beta$ are precisely canceled by Eq.~\eqref{eq:hadamard-R-robin-identity}. The same argument holds in the primed variable. Therefore,
\begin{equation}
 \left.(\partial_z+\beta)H_\beta\right|_{z=0}=\left.(\partial_{z'}+\beta)H_\beta\right|_{z'=0}=0.
\label{eq:hadamard-H-robin}
\end{equation}

Combining Eq.~\eqref{eq:final-two-point} with Proposition~\ref{prop}, we obtain
\begin{equation}
W-H_\beta=S_{\mathrm{un}}\in C^\infty(M\times M).
\label{eq:hadamard-smooth-remainder}
\end{equation}
Therefore, all positive quasifree states constructed in Sec.~\ref{subsec:positive-unstable} share the same singularity structure, independently of their unstable Cauchy covariances. In particular, the unstable sector does not introduce additional singularities, even though its inclusion is necessary to preserve the CCR and prevents the resulting states from being stationary. The explicit identification of the direct and reflected contributions shows that the singularity structure is consistent with Ref.~\cite{CosteriDappiaggiJuarezAubry2026}. A complete microlocal characterization of these states, however, is beyond the scope of the present analysis.

Finally, notice that $H_\beta$ is not itself required to be of positive type. Its unstable contribution contains only the antisymmetric kernel fixed by the CCR, whereas positivity of the complete two-point function requires the additional symmetric covariance $S_{\mathrm{un}}$. The subtraction kernel is also not unique, since a smooth symmetric Robin bisolution can be added without changing its singularity structure or antisymmetric part. Thus, Eq.~\eqref{eq:hadamard-spectral-representative} provides a convenient state-independent representative for the subtraction, rather than a unique prescription for the finite renormalization of local observables.

\section{Conclusion}
\label{sec:conclusion}

We investigated the quantization of a real massless scalar field in four-dimensional half-Minkowski spacetime subject to an attractive Robin boundary condition. The boundary interaction supports a normalizable bound mode, whose dynamics are stable or unstable depending on the tangential momentum. Although the resulting Hamiltonian is indefinite, the total energy remains conserved.

At the classical level, the unstable sector is required for the completeness of the spectral decomposition. At the quantum level, its inclusion is also essential to recover the complete causal propagator and the CCR. In particular, we showed that quantizing only the stable sector---while retaining the unstable modes as classical degrees of freedom---would omit a nonzero contribution to the field commutator, yielding an incomplete causal kernel that fails to reproduce the CCR of the full field.

A consistent quantization can therefore be achieved without assigning a positive-frequency decomposition to the unstable modes. To construct positive quasifree states, we analyzed the symmetric part of the unstable two-point function and established the conditions on its Cauchy covariances. In particular, the uncertainty inequality~\eqref{eq:uncertainty} ensures positivity, while the condition~\eqref{eq:unstable-admissibility} guarantees the smoothness of the unstable contribution. Within the resulting family of states, however, positivity is incompatible with stationarity, so the time-translation symmetry of the background is necessarily broken. Yet, we showed that the associated temporal growth does not contradict energy conservation, since the unstable-sector energy remains finite for each admissible covariance.

Finally, we obtained a closed position-space representation of the complete two-point function, identifying the direct and reflected singularities, including the Robin-dependent logarithmic contributions. From this expression, we identified a state-independent subtraction kernel that satisfies the field equation, the Robin boundary condition, and the complete CCR. Subtracting this kernel leaves only the smooth symmetric contribution of the unstable sector, showing that all states in the constructed family share the same singularity structure, consistent with the Robin--Hadamard state of Refs.~\cite{CosteriEtAl2026,CosteriDappiaggiJuarezAubry2026}. A complete microlocal characterization, however, remains beyond the scope of the present work.

Taken together, our results show that boundary-induced instabilities are compatible with a consistent quantization of the complete field, even though they prevent the construction of stationary positive quasifree states within the family considered. These states share the same local singularity structure, and the smooth differences between their two-point functions allow finite comparisons of local observables through point splitting, without requiring a stationary vacuum. Extending this analysis to curved backgrounds or more general timelike boundaries will require further investigation of both the unstable spectral contributions and the boundary-modified singularity structure. The present model thus provides an explicit setting in which dynamical instabilities can be distinguished from the ultraviolet properties of quantum states.

\begin{acknowledgments}
B.S.F. acknowledges support from the São Paulo Research Foundation (FAPESP, Brazil), Grant 2025/19841-2. J.P.M.P. acknowledges partial support from Conselho Nacional de Desenvolvimento Cient\'ifico e Tecnol\'ogico (CNPq, Brazil) under Grant No.~305194/2025-9 and by Funda\c{c}\~ao de Amparo \`a Pesquisa do Estado de S\~ao Paulo (FAPESP) under Grant No.~2024/00923-6. C.D. acknowledges the support of the Gruppo Nazionale della Fisica Matematica (Indam) and of the INFN, Sezione di Pavia during the realization of this work.

The authors used OpenAI's ChatGPT (GPT-6) to assist with exploratory discussions, text drafting and revision, and \LaTeX\ formatting. All AI-assisted content was thoroughly reviewed and verified by the authors, who take full responsibility for the content of this work.

\end{acknowledgments}

\section*{Data availability}

No data were created or analyzed in this study.


\begin{thebibliography}{99}

  \bibitem{GinouxMurro2022}
  N. Ginoux and S. Murro,
  On the Cauchy problem for Friedrichs systems on globally hyperbolic manifolds with timelike boundary,
  Adv. Differ. Equ. \textbf{27}, 497 (2022).

  \bibitem{Wald1980}
  R. M. Wald,
  Dynamics in nonglobally hyperbolic, static space-times,
  J. Math. Phys. \textbf{21}, 2802 (1980).

  \bibitem{IshibashiWaldII}
  A. Ishibashi and R. M. Wald,
  Dynamics in non-globally-hyperbolic static spacetimes II: General analysis of prescriptions for dynamics,
  Class. Quantum Grav. \textbf{20}, 3815 (2003).

  \bibitem{de2023hidden}
  L. de Souza Campos, C. Dappiaggi, and L. Sinibaldi,
  Hidden freedom in the mode expansion on static spacetimes,
  Gen. Relativ. Gravit. \textbf{55}, 50 (2023).

  \bibitem{DappiaggiFerreiraMarta2018}
  C. Dappiaggi, H. R. C. Ferreira, and A. Marta,
  Ground states of a Klein-Gordon field with Robin boundary conditions in global anti-de Sitter spacetime,
  Phys. Rev. D \textbf{98}, 025005 (2018).

  \bibitem{Morley2020Quantum}
  T. Morley, P. Taylor, and E. Winstanley,
  Quantum field theory on global anti-de Sitter space-time with Robin boundary conditions,
  Class. Quantum Grav. \textbf{38}, 035009 (2021).

  \bibitem{PitelliEtAl2025}
  J. P. M. Pitelli, B. S. Felipe, C. Dappiaggi, and E. Winstanley,
  Quantum superposition of boundary conditions in PAdS$_2$,
  Phys. Rev. D \textbf{111}, 125011 (2025).

  \bibitem{PhysRevD.89.025020}
  T. R. Govindarajan and V. P. Nair,
  Quantum field theories with boundaries and novel instabilities,
  Phys. Rev. D \textbf{89}, 025020 (2014).

  \bibitem{Asorey2015Edge}
  M. Asorey, A. P. Balachandran, and J. M. P\'erez-Pardo,
  Edge states at phase boundaries and their stability,
  Rev. Math. Phys. \textbf{28}, 1650020 (2016).

  \bibitem{DappiaggiFerreira2016}
  C. Dappiaggi and H. R. C. Ferreira,
  Hadamard states for a scalar field in anti-de Sitter spacetime with arbitrary boundary conditions,
  Phys. Rev. D \textbf{94}, 125016 (2016).

  \bibitem{FelipePitelli2024}
  B. S. Felipe and J. P. M. Pitelli,
  Quantum approach to bound states in field theory,
  Phys. Rev. D \textbf{109}, 105013 (2024).

  \bibitem{KayWald1991}
  B. S. Kay and R. M. Wald,
  Theorems on the uniqueness and thermal properties of stationary, nonsingular, quasifree states on spacetimes with a bifurcate Killing horizon,
  Phys. Rep. \textbf{207}, 49 (1991).

  \bibitem{Radzikowski1996}
  M. J. Radzikowski,
  Micro-local approach to the Hadamard condition in quantum field theory on curved space-time,
  Commun. Math. Phys. \textbf{179}, 529 (1996).

  \bibitem{Bondurant_2005}
  J. D. Bondurant and S. A. Fulling,
  The Dirichlet-to-Robin transform,
  J. Phys. A: Math. Gen. \textbf{38}, 1505 (2005).

  \bibitem{CosteriEtAl2026}
  B. Costeri, C. Dappiaggi, B. A. Ju\'arez-Aubry, and R. D. Singh,
  The Hadamard parametrix on half-Minkowski with Robin boundary conditions: Fundamental solutions and Hadamard states,
  Ann. Henri Poincar\'e (2026).

  \bibitem{CosteriDappiaggiJuarezAubry2026}
  B. Costeri, C. Dappiaggi, and B. A. Ju\'arez-Aubry,
  The Hadamard parametrix on globally hyperbolic spacetimes with Robin boundary conditions: Fundamental solutions and Hadamard states (2026).

  \bibitem{BayonaBraga2007}
  C. A. Ballon Bayona and N. R. F. Braga,
  Anti-de Sitter boundary in Poincar\'e coordinates,
  Gen. Relativ. Gravit. \textbf{39}, 1367 (2007).

  \bibitem{Saharian2004}
  A. A. Saharian,
  Energy-momentum tensor for a scalar field on manifolds with boundaries,
  Phys. Rev. D \textbf{69}, 085005 (2004).

  \bibitem{Morley2023Renormalized}
  T. Morley, S. Namasivayam, and E. Winstanley,
  Renormalized stress-energy tensor on global anti-de Sitter space-time with Robin boundary conditions,
  Gen. Relativ. Gravit. \textbf{56}, 38 (2024).

  \bibitem{Lima2013}
  W. C. C. Lima,
  Quantization of unstable linear scalar fields in static spacetimes,
  Phys. Rev. D \textbf{88}, 124005 (2013).

  \bibitem{Khavkine2015}
  I. Khavkine and V. Moretti,
  Algebraic QFT in curved spacetime and quasifree Hadamard states: An introduction, in \textit{Advances in Algebraic Quantum Field Theory} (Springer, Cham, 2015).

  \bibitem{Kay2006Quantum}
  B. S. Kay,
  Quantum field theory in curved spacetime, in \textit{Encyclopedia of Mathematical Physics} (Academic Press, Oxford, 2025).

  \bibitem{Hack2016AlgebraicQF}
  T.-P. Hack,
  Algebraic quantum field theory on curved spacetimes,
  in \textit{Cosmological Applications of Algebraic Quantum Field Theory in Curved Spacetimes},
  SpringerBriefs in Mathematical Physics (Springer, Cham, 2016).

  \bibitem{Dappiaggi2019}
  C. Dappiaggi, N. Drago, and H. Ferreira,
  Fundamental solutions for the wave operator on static Lorentzian manifolds with timelike boundary,
  Lett. Math. Phys. \textbf{109}, 2157 (2019).

  \bibitem{Robertson1929}
  H. P. Robertson,
  The uncertainty principle,
  Phys. Rev. \textbf{34}, 163 (1929).

  \bibitem{gradshteyn2014table}
  I. S. Gradshteyn and I. M. Ryzhik,
  \textit{Table of Integrals, Series, and Products}  (Academic Press, 2014).

  \bibitem{DLMF}
  D. W. Lozier,
  NIST digital library of mathematical functions,
  Ann. Math. Artif. Intell. \textbf{38}, 105 (2003).

\end{thebibliography}
\end{document}